\documentclass[%
reprint,amsmath,amssymb,aps,prl,superscriptaddress,longbibliography,
floatfix,
]{revtex4-2}

\usepackage{graphicx}
\usepackage{dcolumn}
\usepackage{bm}
\usepackage{epstopdf}
\usepackage{float}
\usepackage{braket}
\usepackage{dsfont}
\usepackage{amsmath}
\usepackage{siunitx} 
\usepackage[italicdiff]{physics}
\usepackage[T1]{fontenc}
\usepackage[dvipsnames]{xcolor}
\usepackage{color,soul}
\usepackage{lipsum}
\usepackage[colorlinks=true, allcolors=blue]{hyperref}

\begin{document}

\title{Stabilizing an individual charge fluctuator in a Si/SiGe quantum dot}

\author{Feiyang Ye}
\thanks{These authors contributed equally.}

\author{Ammar Ellaboudy}
\thanks{These authors contributed equally.}

\affiliation{Department of Physics and Astronomy, University of Rochester, Rochester, New York 14627, USA}

\author{John M. Nichol}
\email{john.nichol@rochester.edu}
\affiliation{Department of Physics and Astronomy, University of Rochester, Rochester, New York 14627, USA}

\begin{abstract}
Charge noise is a major obstacle to improved gate fidelities in silicon spin qubits.  Numerous methods exist to mitigate charge noise, including improving device fabrication, dynamical decoupling, and real-time parameter estimation. In this work, we demonstrate a new class of techniques to mitigate charge noise in semiconductor quantum dots by controlling the noise sources themselves. Using two different classical feedback methods, we stabilize an individual charged two-level fluctuator in a Si/SiGe quantum dot by exploiting sensitive gate-voltage dependence of the switching times. These control methods reduce the low-frequency component of the noise power spectrum by an order of magnitude. These techniques also enable stabilizing the fluctuator in either of its states. In the future, such techniques may enable improved coherence times in quantum-dot spin qubits. 
\end{abstract}

\pacs{}

\maketitle

\section{Introduction}

Spin qubits based on gate-defined semiconductor quantum dots are excellent qubits due to their long coherence times, small size, and compatibility with advanced semiconductor manufacturing techniques~\cite{burkard2023semiconductor}. Most gate-defined quantum dots rely on precise control of the electrostatic confinement potential to achieve initialization, readout, as well as single- and multi-qubit gates. Thus, random electrical noise in the device that disturbs the confinement potentials, or charge noise, is a major obstacle in improving spin qubit fidelity. While not completely understood, the microscopic description of charge noise likely involves an ensemble of two-level fluctuators (TLFs) in the host material, which result in a 1/$f$-like noise power spectrum~\cite{Paladino2014,burkard2023semiconductor,Freeman2016,yoneda_quantum-dot_2018,mi_landau-zener_2018,connors2019low,struck2020low,petit_spin_2018,rudolph2019long,kranz2020exploiting,elsayed2022low,holman20213d,connors2022charge,paquelet2023reducing,ye2024characterization}. 

Potential options for mitigating charge noise involve design choices including the width~\cite{paquelet2023reducing} or depth~\cite{kranz2020exploiting} of the quantum well, the morphology of any capping layers~\cite{paquelet2023reducing}, or the thickness of the gate dielectric~\cite{connors2019low}. Dynamical-decoupling protocols can filter the low-frequency noise of the environment, increasing the coherence time of the qubit \cite{yoneda_quantum-dot_2018,medford2012scaling,bylander_noise_2011,connors2022charge}. Real-time feedback of sensor~\cite{nakajima2021real} or qubit~\cite{Shulman2014,park2025passive,berritta2024real} parameters can also reduce the effect of charge fluctuations.  Another promising technique involves stabilizing the fluctuators themselves. This strategy has been applied effectively in the case of hyperfine fluctuations in spin-qubit systems~\cite{Bluhm2010,Nichol2017}. The possibility of coherent control of electrically active two-level systems has been established in superconducting quantum systems~\cite{lisenfeld2010measuring}.  Additionally, two-level systems, just like qubits, can also be dynamically decoupled through a variety of mechanisms~\cite{matityahu2019dynamical, niepce2021stability}. 

\begin{figure}[hb]
{\includegraphics[width= 0.46\textwidth]{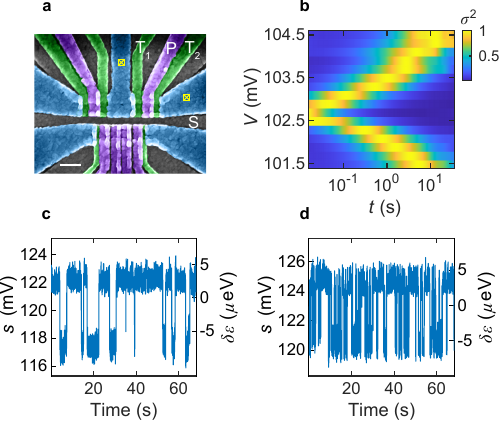}}
\caption{
\textbf{Experimental setup.}
\textbf{a} The experiments reported here use the upper right quantum dot in a multi-dot device. The white scale bar is $100~\si{nm}$. 
\textbf{b} Normalized Allan variance of the TLF signal at different gate voltages $V$ applied to the screening gate S. The Allan variance of a random telegraph signal has a peak at a time lag $t$ approximately equal to the TLF average switching time \cite{PRINCIPATO200775}.
For each value of $V$, we also adjust the plunger-gate voltage to keep the quantum-dot chemical potential fixed.  The TLF switching time depends sensitively on gate voltages.
\textbf{c} Example time series for the slow tuning.
\textbf{d} Example time series for the fast tuning.
The TLF causes about $10~\si{\mu eV}$ chemical potential fluctuations.
}
\label{fig:device}
\end{figure}

In this work, we report two feedback methods to stabilize an electrically active TLF in a Si/SiGe quantum dot. Both techniques leverage the sensitive voltage dependence of the TLF switching times in quantum dots~\cite{ye2024characterization}, and both methods involve monitoring the state of the TLF in real time. When the TLF switches its state, we change the gate voltages to a configuration where the TLF switching time decreases. In the first open-loop method, we wait a fixed amount of time at this fast tuning, before returning to the original slow tuning. In the second closed-loop method, we monitor the TLF state in the fast tuning and return to the slow tuning after it has switched back to the desired state.  Both methods reduce the low-frequency fluctuations of the TLF by nearly an order of magnitude.  We also demonstrate that feedback can be used to stabilize the TLF in either of its states, and the performance of both agrees with numerical simulations. In the future, these methods could be used to suppress charge noise in spin-qubit devices, improving their coherence times. 

\begin{figure*}[ht]
{\includegraphics[width=\textwidth]{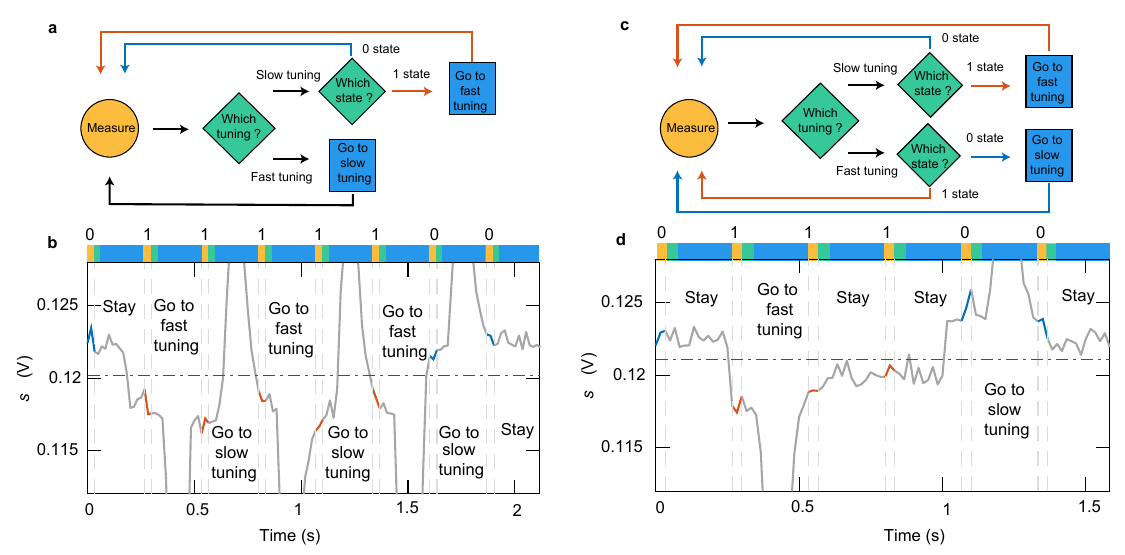}}
\caption{ 
\textbf{Feedback methods.} 
\textbf{a} In the open-loop control method, we monitor the TLF state in the slow tuning.  If the TLF is measured in the 1 state, we change to the fast tuning for a set amount of time, and then return to the slow tuning.
\textbf{b} Actual time series recorded during the open-loop process, with decisions indicated.  The gray trace is the actual measured signal. The large deviations in the signal are the result of the voltage pulses during the tuning change. The colored segments indicate the state of the TLF determined during each measurement segment, where the signal threshold is indicated by the horizontal dashed line. 
\textbf{c} In the closed-loop control method, we also monitor the state of the TLF in the fast tuning, and we return to the slow tuning once the TLF is found in the 0 state. 
\textbf{d} Actual time series recorded during the closed-loop process, with decisions indicated. 
}
\label{fig:methods}
\end{figure*}

\section{Experimental Setup}
The experiments described here involve a single quantum dot in a Si/SiGe heterostructure with an 8-nm-wide quantum well 50~nm below the surface of the semiconductor (Fig.~\ref{fig:device}a). A 15-nm-thick aluminum oxide gate dielectric is deposited on the chip by atomic layer deposition. The device is fabricated with an overlapping gate architecture and is configured for radiofrequency reflectometry~\cite{connors2020rapid}.  We operate the device at a temperature of 10~mK in a dilution refrigerator.  The other dots in the device are either tuned deep in the Coulomb blockade regime or the accumulation regime with many electrons. 

We tune the quantum dot to the Coulomb blockade regime and fix its chemical potential on the side of a transport peak, where the conductivity of the dot is maximally sensitive to chemical potential fluctuations. We measure changes in the dot conductance via radiofrequency reflectometry. As reported previously, we find pronounced random telegraph noise, with the TLF switching between two states, which we label as ``0'' and ``1''. Based on measurements of the TLF occupation probability, we identify the 0 state as the ground state. Moreover, we observe that the characteristic switching time depends sensitively on gate voltages~\cite{ye2024characterization} (Fig.~\ref{fig:device}b).

In the remainder of this work, we will use two different voltage settings, or tunings, which feature different TLF switching times. We refer to the first as the ``slow tuning'', where the switching time is on the order of seconds or slower (Fig.~\ref{fig:device}c). We also use a ``fast tuning'',  where the switching time is of order seconds or faster (Fig.~\ref{fig:device}d). These two tunings differ by voltages applied to both the plunger gate and screening gate (Fig.~\ref{fig:device}b), such that the chemical potential of the quantum dot is approximately the same in both tunings~\cite{ye2024characterization}. In Figs.~\ref{fig:device}c and d, the chemical potential fluctuations $\delta \varepsilon$ are converted from variations in the reflectometry signals $\delta s$ via the relation $\delta \varepsilon = a \delta s/(ds/dV_P)$ \cite{connors2019low}, with $a=0.053~\si{eV/V}$ the lever arm and $ds/dV_P$ the sensor sensitivity at the configuration for experiments.

\begin{figure*}[ht!]
\centering
{\includegraphics[width=\textwidth]{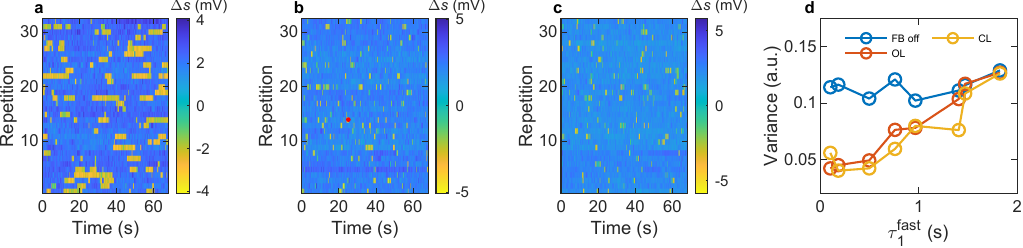}}
\caption{
\textbf{Stabilizing the TLF in the 0 state.}
\textbf{a} Representative measurements of the TLF without any feedback. The vertical axis represents different repetitions. The blue color corresponds to the 0 state, and the yellow color corresponds to the 1 state. The colorbar $\Delta s$ shows the difference between the measured signal and the threshold.
\textbf{b} and \textbf{c} show measurements of the same TLF with open-loop (OL) and closed-loop (CL) feedback, respectively. The red point in \textbf{b} marks an error in our readout, because the gate-voltage change overlaps with the readout window resulting in a large signal deviation from the threshold.
\textbf{d} Measured variance of the TLF signal vs $\tau_1^{fast}$.
}
\label{fig:upstate}
\end{figure*}

In the following, we will refer to the mean time spent in the 0 state before a transition to the 1 state in the fast (slow) tunings without feedback as $\tau^{fast}_{0}$ ($\tau^{slow}_{0}$). Likewise, we will denote the mean time spent in the 1 state before a switch to the 0 state without feedback as $\tau^{fast}_{1}$ ($\tau^{slow}_{1}$).
With feedback on, we will denote the effective mean life time spent in the 0 (1) state as $\tau^{feedback}_0$ ($\tau^{feedback}_1$).
For most of the experiments described below, we will stabilize the TLF in the 0 state. Thus, we will expect $\tau_1^{fast}<\tau^{feedback}_{1}<\tau_1^{slow}$ with feedback, as the TLF spends less time in its excited state.

\section{Open-loop control}
The primary idea of our open-loop feedback method is that the switching times of the TLF decrease significantly in the fast tuning, compared to the slow tuning. If we wish to stabilize the TLF in the 0 state, and if we find it to be in the 1 state in the slow tuning, we can significantly increase the probability of finding it back in the 0 state by waiting for a time on the order of $\tau^{fast}_{1}\ll \tau^{slow}_{1}$ in the fast tuning.  In this procedure, we divide the total feedback cycle, which takes $\Delta T=8/30$ s, into two parts: measurement and feedback (Figs.~\ref{fig:methods}a-b). The measurement segment takes $\Delta T/8$. We use the reflectometry signal acquired during this time to determine the TLF state. In the feedback segment, which lasts $7 \Delta T /8$, we change the gate voltages if required. Specifically, if we find the TLF in the 1 state during the measurement window, we change to the fast tuning and then return to the slow tuning during the next feedback segment. Due to the low-pass filters installed in our setup, the time to change the gate voltages occupies a significant fraction of $7\Delta T/8$ and takes a few hundred milliseconds (Fig.~\ref{fig:methods}b). If, during the measurement window, we find the TLF in the 0 state, we do not change the tuning. After the feedback segment, the entire cycle repeats.  In these experiments, the length of the feedback cycle is primarily constrained by the length of time it takes to change the gate voltages. As discussed further below, significant improvements can be made to this setup to increase its speed.

Figure~\ref{fig:upstate}a shows representative time-series data for the TLF we measure without any feedback. Figure~\ref{fig:upstate}b shows time-series data for the same TLF with open-loop feedback. Visual inspection confirms effective stabilization of the 0 state. To assess the performance of this method, we perform 32 repetitions of 256 feedback cycles, acquiring a time series of length $256 \times \Delta T \approx 68$~s for each repetition. To avoid errors in our state assignment due to drifts in the tuning, we update the 0/1 signal threshold every four repetitions. 

To quantify the performance of the feedback, we threshold the resulting time series data using a Gaussian Mixture Model to determine the state of the TLF as a function of time. We plot the variance of the resulting time series for the case with no feedback and the cases with feedback, and we compare them in Fig.~\ref{fig:upstate}d. As expected, the variance decreases with feedback applied. The variance also decreases together with $\tau_1^{fast}$, because the feedback performs better with smaller $\tau_1^{fast}$.  Below, we discuss the power spectral density of the TLF in the presence of feedback.

\section{Closed-loop control}
The closed-loop feedback method is similar to the open-loop method, except that we monitor the state of the TLF in the fast tuning, and we return to the slow tuning only when the TLF has switched back to the desired state (Figs.~\ref{fig:methods}c,d). As before, we stabilize the TLF in the 0 state in the slow tuning. 

Figure~\ref{fig:upstate}c shows the time series of the TLF signal when the feedback is applied. Compared with Fig.~\ref{fig:upstate}a again shows significantly reduced time in the 1 state. Figure~\ref{fig:upstate}d plots the variance of the TLF signal vs $\tau_1^{fast}$, illustrating the performance of closed-loop method. As expected, the variance shows the same trend with $\tau_1^{fast}$ as the open-loop method. However, the variance is only marginally lower than the open-loop method. As we discuss below in the modeling section, this has primarily to do with the relatively large value of the total feedback cycle $\Delta T$.


\begin{figure}[t]
{\includegraphics[width= 0.5\textwidth]{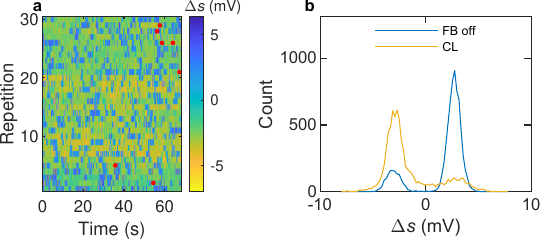}}
\caption{\textbf{Stabilizing the TLF in the 1 state.}
\textbf{a} Time series while the closed-loop feedback is on. 
\textbf{b} Histogram of sample time series with feedback on and off.
}
\label{fig:downstate}
\end{figure}
 
To demonstrate the utility of the closed-loop method, we also stabilize the TLF in its excited state (the 1 state). Figure~\ref{fig:downstate}a shows the time series data for the case with the feedback on.  
With the feedback applied, the TLF spends more time in its excited state (Fig.~\ref{fig:downstate}b).


\section{Modeling}
We model our results in two ways. The first is based on numerical simulations.
In these simulations, we treat the TLF as a classical random telegraph fluctuator and perform Monte Carlo simulations of the TLF time traces for different feedback methods.
We calculate the transition probability at each time step $dt$ as $dt/\tau$, where the state- and tuning-dependent switching time $\tau$ is extracted from the experiments without feedback.
We assume that the TLF starts with the 1 state in the slow tuning and simulate the different feedback procedures shown in Fig.~\ref{fig:methods}.
We simulate the TLF with time step $dt=\Delta T/16 = 1/60~\mathrm{s}$, and perform $64$ repetitions of $256$ feedback cycles to mimic our experiments. The simulations without feedback are fixed in the slow tuning.

\begin{figure}[t]
{\includegraphics[width= 0.5\textwidth]{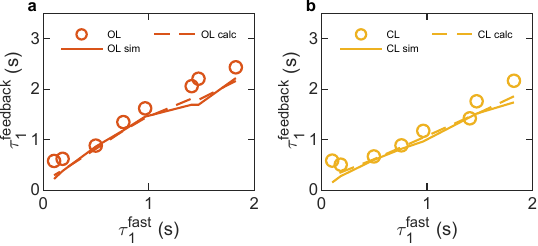}}
\caption{\textbf{Modeling $\tau_1^{feedback}$.}
\textbf{a} Measured, simulated, and calculated values of $\tau_1^{feedback}$ vs $\tau_1^{fast}$ during open-loop stabilization of the 0 state. Here ``calculated'' refers to the semi-analytical approach described in the text. 
\textbf{b}  Measured, simulated, and calculated values of $\tau_1^{feedback}$ vs $\tau_1^{fast}$ during closed-loop feedback of the 0 state.}
\label{fig:taumodeling}
\end{figure}

We first compare the measured and simulated values of $\tau_1^{feedback}$. We determine the experimental values of $\tau_1^{feedback}$ as the average time spent in the 1 state based on the TLF signal. Figure~\ref{fig:taumodeling}a shows the measured and simulated values for open-loop feedback, and Fig.~\ref{fig:taumodeling}b shows the measured and simulated values for closed-loop feedback. As $\tau_{1}^{fast}$ decreases, the feedback performs better, and the TLF spends less time in the 1 state, and $\tau_1^{feedback}$ becomes smaller. In all cases, the simulations agree quantitatively with the measurements. 

\begin{figure*}[t]
{\includegraphics[width= 0.9\textwidth]{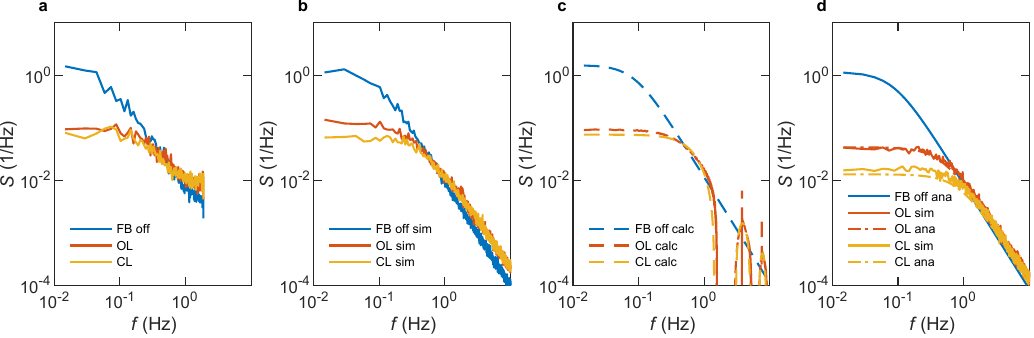}}
\caption{\textbf{Modeling the TLF power spectral density.}
\textbf{a} Power spectra of the measured time series without feedback and during open- and closed-loop feedback.
\textbf{b} Numerical simulations of the power spectra. The high-frequency noise level slightly increases with feedback.
\textbf{c} Calculated power spectra based on the model described in the text.
\textbf{d} Simulated power spectra, for the case that the feedback time can be reduced to $\Delta T=2/60~\mathrm{s}$. The ``FB off ana'' line  is a Lorentzian power spectrum assuming $\tau_0=\tau_0^{slow}$ and $\tau_1=\tau_1^{slow}$, the ``CL ana'' line is a Lorentzian with $\tau_0=\tau_0^{slow}$ and $\tau_1=\tau_1^{fast}$ assuming $\Delta T$ is infinitely small, and the ``OL ana'' line is a Lorentzian with $\tau_0=\tau_0^{slow}$ and $\tau_1=2/(1/\tau_1^{fast}+1/\tau_1^{slow})$.
}
\label{fig:psdmodeling}
\end{figure*}

Figure~\ref{fig:psdmodeling}a shows the power spectral densities of the TLF signal without feedback and with open- and closed-loop feedback for the case when $\tau_1^{fast}=0.18$ s.
For frequencies below about 1 Hz, the power spectral density of fluctuations associated with this TLF decreases by more than an order of magnitude with feedback.
Figure~\ref{fig:psdmodeling}b shows the simulated power spectral densities for the three cases, which agree quantitatively with the data. 

The second method we use to understand our results is a semi-analytical approach. Let us assume that at $t=0$, the TLF is in the $1$ state. Following Ref.~\cite{tlfpage}, we define $P_{11}(t)$ as the probability that the TLF undergoes an even number of transitions in time $t$. Likewise, we define $P_{10}(t)$ as the probability to undergo an odd number of transitions in $t$. Thus, 
\begin{align}
P_{11}(t+dt)=P_{10}(t)\frac{dt}{\tau_0}+P_{11}(t)\left(1-\frac{dt}{\tau_1} \right),
\end{align} 
where $\tau_0$ is the mean time spent in the $0$ state, and $\tau_1$ is the mean time spent in the $1$ state. Rearranging, and using the fact that $P_{11}(t)+P_{10}(t)=1$,
\begin{align}
\frac{dP_{11}(t)}{dt}+P_{11}(t)\left( \frac{1}{\tau_0}+\frac{1}{\tau_1}\right)=\frac{1}{\tau_0}.
\end{align}
A straightforward approach to solving this equation is to use an integrating factor $\alpha(t)=\exp\left[ \int_0^t \left( \frac{1}{\tau_0}+\frac{1}{\tau_1}\right) dt'\right]$. The solution is formally written as 
\begin{align}
P_{11}(t)=\frac{\int_0^t \frac{\alpha(t')}{\tau_0} dt'}{\alpha(t)}+\frac{\alpha(0)}{\alpha(t)},
\label{eq:p11}
\end{align}
where $P_{11}(0)=1$.

Without feedback, $\tau_0$ and $\tau_1$ are constants, and the solution $P_{11}(t)$ is an exponential function of time with time constant $(1/\tau_0+1/\tau_1)^{-1}$. To compute the full autocorrelation function, we multiply $P_{11}(t)$ by the time-averaged probability to find the TLF in the 1 state $P_1$. In the case of fixed switching times, $P_1= \tau_1/(\tau_0+\tau_1)$. Fourier-transforming the autocorrelation function yields the characteristic Lorentzian power spectrum for an individual TLF:
\begin{equation}
    S(f,\tau_0,\tau_1) = \frac{4A^2}{(\tau_0 + \tau_1) \qty[(1/\tau_0 + 1/\tau_1)^2 + (2\pi f)^2]},
\label{eq:psd}
\end{equation}
with $A$ the TLF peak-to-peak amplitude and $f$ the frequency.

To use this framework to model the effect of feedback applied, we assume that $\tau_1$ depends on time, but that $\tau_0$ remains fixed, because the tuning reverts to the slow tuning once the TLF switches back to the $0$ state. We assume that for the case of open-loop feedback $\tau_1$ has the following time dependence
\begin{align}
    \tau_1(t)=
    \begin{cases}
    \tau_1^{slow} \text{ for } 0\leq t<\Delta T\\
    \tau_1^{fast} \text{ for } \Delta T \leq t<2 \Delta T\\
    \tau_1^{slow} \text{ for } 2\Delta T \leq t<3 \Delta T\\
    \tau_1^{fast} \text{ for } 3\Delta T \leq t<4 \Delta T\\
    \tau_1^{slow} \text{ for } 4\Delta T \leq t<5 \Delta T\\
    \cdots \\
    \end{cases}.
\end{align}
Here the first period in the slow tuning represents the delay in identifying the state of the TLF, and then the following periods represent the fast-slow repetition of the open-loop feedback. For the case of closed-loop feedback, we use 
\begin{align}
    \tau_1(t)=
    \begin{cases}
    \tau_1^{slow} \text{ for } 0\leq t<\Delta T\\
    \tau_1^{fast} \text{ for } \Delta T \leq t<\infty\\
    \end{cases}.
\end{align}
With these forms of $\tau_1(t)$, we numerically compute $\tau_1^{feedback}$ by fitting $P_{11}(t)$ to an exponential with time-constant $\tau$. We extract the effective time spent in the $1$ state as $\tau_1^{feedback}=(1/\tau-1/\tau_0)^{-1}$, and we compute $P_1=\tau_1^{feedback}/(\tau_0+\tau_1^{feedback})$. For each of the experimental configurations, we plot these extracted times in Fig.~\ref{fig:taumodeling}, and they agree with the measured and simulated times. 

Figure~\ref{fig:psdmodeling}c shows the power spectral densities computed as the cosine transform of the autocorrelation function $P_1 \times P_{11}(t)$ for the case of no feedback, open-loop feedback, and closed-loop feedback. These calculations reproduce the main features of the data and simulations. The structure at high frequencies is likely due to the periodic modulation of $\tau_1$. More advanced modeling would include the fact that the phase of this modulation changes because the initial time lag is random, and the structure would likely be smoothed out. 

We gain insight into the nature of this calculation by considering the limit of fast feedback cycles, where $\Delta T \to 0$. 
Let $t = 2n \Delta T$ with $n$ an integer. For the open-loop method, the integrating factor $\alpha(t)$ reads
\begin{align}
 \alpha(t) & = \exp\qty[\int_0^t (1/\tau_0 + 1/\tau_1) dt' ] \notag \\
   &= \exp\qty[ t/\tau_0 + \int_0^{\Delta T} 1/\tau_1 dt' + \int_{\Delta T}^{2 \Delta T} 1/\tau_1 dt' + ... ] \notag \\
   &=\exp \qty[t/\tau_0 + n \Delta T (1/\tau_1^{slow} + 1/\tau_1^{fast})] \notag \\
   &= \exp \qty[ \qty( 1/\tau_0 + (1/\tau_1^{slow} + 1/\tau_1^{fast})/2 ) t].
\end{align}
Hence, in the fast-loop limit, $P_{11}(t)$ in Eq.~\eqref{eq:p11} is an exponential function of time, and the effective switching rate out of the 1 state is the average of the fast and slow rates $(1/\tau_1^{slow}+1/\tau_1^{fast})/2$.
For the closed-loop method, $\tau_1$ approaches $\tau_1^{fast}$ when $\Delta T \to 0$ such that $\tau_1^{feedback}=\tau_1^{fast}$.

Figure~\ref{fig:psdmodeling}d shows simulations and calculations for both open- and closed-loop feedback when the feedback interval is reduced to $\Delta T=2/60$~s. In this case, the closed-loop scheme performs noticeably better than the open-loop scheme. We also plot Lorentzian power spectra assuming $\tau_1^{feedback}=\tau_1^{fast}$, and $\tau_1^{feedback}=2/(1/\tau_1^{fast}+1/\tau_1^{slow})$, which agree with the closed- and open-loop predictions, respectively. This agreement reflects the fact that when the loop time is small, the open-loop feedback changes the effective switching rate of the 1 state $1/\tau_{1}^{feedback}$ to the average of the fast and slow rates, while the closed-loop feedback changes the effective switching rate to the fast rate. 

\section{Conclusion}
Low-frequency charge noise is a major challenge affecting spin qubit gate fidelities. In this work, we have demonstrated a class of techniques that mitigate charge noise by directly controlling the sources of noise themselves. A number of improvements could be made to our setup to increase its effectiveness. The longest elements of our feedback cycle, which currently lasts 267 ms, involve data transfer between the data acquisition card and our software, and the voltage pulses, which pass through low-pass filters. Instead, with a field-programmable gate array and wiring compatible with fast voltage pulses, the cycle time could likely be reduced to the millisecond level or below, significantly increasing the bandwidth and performance of our techniques. 

Recent theoretical work has shown that individual TLFs can indeed have a significant impact on qubit performance~\cite{mehmandoost2024decoherence}, and that removing such fluctuators can enhance qubit coherence. Thus, the techniques we have discussed to stabilize individual TLFs may help to improve spin-qubit coherence.  While we have demonstrated the use of feedback to stabilize a TLF coupled to a charge sensor, applying our techniques to spin qubits will depend critically on the ability to resolve individual voltage-dependent TLFs in real time in spin-qubit devices, likely using the qubits themselves. In prior work, we have found that  TLFs in our devices frequently appear to depend on gate voltages~\cite{ye2024characterization}. Recent work has also demonstrated the possibility of real-time measurements of charge noise using spin qubits~\cite{park2025passive}. While not all TLFs will have a large enough effect to enable rapid real-time measurement using qubits, those TLFs that are large enough for real-time measurement will likely have the most significant effects on spin qubit performance and will thus be the most important to stabilize. In addition, we have demonstrated this technique with a single TLF, but our technique in principle can be applied to multiple TLFs provided that they can be individually resolved and have voltage-dependent switching times. Advanced signal processing techniques, such as hidden Markov modeling~\cite{Albrecht2023}, could potentially be used to distinguish the states of TLFs where simple threshold techniques do not suffice. In addition to reducing the variance of electrical noise, our feedback protocol could potentially assist post-selection strategies, by increasing the number of instances when the environment of a qubit has the desired configuration.


\section{Data Availability}
The processed data that support the findings of this study are available in Ref.~\cite{zenodo}. The raw data are available from the corresponding author
upon reasonable request.

\section{Acknowledgments}
We thank Lisa F. Edge of HRL Laboratories, LLC for the epitaxial growth of the SiGe material and Elliot J. Connors for device fabrication.
This work was sponsored by the Army Research Office through Grant No. W911NF-23-1-0115 and the Air Force Office of Scientific Research through Grant No. FA9550-23-1-0710. The views and conclusions contained in this document are those of the authors and should not be interpreted as representing the official policies, either expressed or implied, of the Army Research Office or the U.S. Government. The U.S. Government is authorized to reproduce and distribute reprints for Government purposes notwithstanding any copyright notation herein.

\section{Author Contributions}
F.Y., A.E., and J.M.N. formulated the experiment and carried out the measurements; all authors wrote the paper; J.M.N. supervised the research.

%

\end{document}